# Anisotropic proximity-induced superconductivity and edge supercurrent in Kagome metal, $K_{1-x}V_3Sb_5$

Yaojia Wang[1,2*], Shuo-Ying Yang[1], Pranava K. Sivakumar[1], Brenden R. Ortiz[3], Samuel M.L. Teicher[3], Heng Wu[1,2], Abhay K. Srivastava[1], Chirag Garg[1], Defa Liu[1], Stuart S. P. Parkin[1], Eric S. Toberer[4], Tyrel McQueen[5], Stephen D. Wilson[3], Mazhar N. Ali[1,2*]

[1]Max Planck Institute of Microstructure Physics, Halle, Saxony-Anhalt, 06108, Germany

[2]Kavli Institute of Nanoscience, Delft University of Technology, Delft, the Netherlands

[3]Materials Department, University of California Santa Barbara, Santa Barbara, CA, 93106, USA

[4]Colorado School of Mines, Goldon, Colorado 80401, USA

[5]Johns Hopkins University, Baltimore, Maryland 21218, USA

*Corresponding author: whyjwang@gmail.com, maz@berkeley.edu

**Materials with transition metals in triangular lattices are of great interest for their potential combination of strong correlation, exotic magnetism and electronic topology. Kagome nets are of particular importance since the discovery of geometrically frustrated magnetism and topological band structures in crystals like Herbertsmithite and $Fe_3Sn_2$, respectively. $KV_3Sb_5$ was discovered to be a layered topological metal with a Kagome net of vanadium. Here, we fabricated Josephson Junctions (JJ) of $K_{1-x}V_3Sb_5$ and induced superconductivity over long junction lengths. Through magnetoresistance and current vs. phase measurements, we observed magnetic field sweeping direction dependent magnetoresistance, and an anisotropic interference pattern with a Fraunhofer pattern for in-plane magnetic field, but a suppression of critical current for out-of-plane magnetic field. These results indicate an anisotropic internal magnetic field in $K_{1-x}V_3Sb_5$ which influences the superconducting coupling in the junction, possibly giving rise to spin-triplet superconductivity. In addition, the observation of long-lived fast oscillations shows evidence of spatially localized conducting channels arising from edge states. These observations pave the way for studying unconventional superconductivity and Josephson device based on Kagome metals with electron correlation and topology.**

## Introduction

The Kagome lattice, formed by corner-sharing triangles of atoms, is an important structure type that is proximate to the honeycomb (like graphene), hosting topological band structures, electron correlation, and geometrical frustration(*1–3*). It is an ideal platform for investigating many exotic electronic behaviors such as the quantum spin liquid state, unconventional superconductivity, and Dirac/Weyl/Nodal line semimetal behavior(*3–6*). Recently, the family of Kagome metals $AV_3Sb_5$ (A=K, Cs, Rb)(*7*) has attracted substantial attention due to the observation of diverse physical properties including charge-density waves (CDW)(*8–13*), superconductivity(*14–17*), a giant anomalous Hall effect (AHE)(*18*, *19*) and topological states(*14*, *18*, *20*, *21*).

Many studies have been performed to explore and understand the nature of these properties and electronic states in the $AV_3Sb_5$ quantum material family. While no long-range magnetic order has been observed through inelastic neutron scattering(*7*), muon spin relaxation experiments have shown signals of time-reversal symmetry (TRS) breaking with weak internal local magnetic field below CDW transition in $KV_3Sb_5$ and $CsV_3Sb_5$(*22–24*). This aligned well with the observed giant extrinsic skew scattering AHE below the CDW transition(*18*, *19*). Also, chiral charge ordering with magnetic field tunability and charge order states breaking rotation-symmetry were detected in $CsV_3Sb_5$ and $KV_3Sb_5$(*9*, *11*, *12*, *25–27*), and possible unconventional superconductivity in $AV_3Sb_5$ was also discussed(*11*, *12*, *22*, *28*). Theoretically, several ground states of $AV_3Sb_5$ were proposed including a star-of-David structure, Tri-hexagonal structures, and TRS breaking states such as a chiral flux state, charge/spin bond orders, spin density waves, etc(*29–34*). However, differing reports on the presence or absence of these order states and the nature of the superconductivity(*25*, *35–37*) has made understanding the magnetism and superconductivity in the $AV_3Sb_5$ family difficult, and there is a lot of space to excavate the underlying physics.

In addition to looking for intrinsic superconductivity, utilizing the proximity effect is another route to investigate the superconducting and electronic properties of materials. The quantum material Josephson junction (JJ), where two superconductors (SC) are coupled via a bridge of non-superconducting (non-SC) quantum material, is an effective structure to induce superconductivity. In a JJ, proximity-induced Cooper pairs are very sensitive to the barrier properties, and the resultant interference effects based on the coupling of the SC wavefunctions can reveal inherent physical properties of the barriers, including topological states and magnetism(*38–40*). On the other hand, the Josephson junction is an important structure to explore intriguing physical properties, such as generating spin-triplet Cooper pairs in JJs of magnetic barriers(*41–48*) as well as exploring possible topological superconductivity through superconductivity of a topological state(*49*, *50*). Thus, Josephson junctions with topological Kagome metal barriers are a great tool to explore intriguing superconducting properties and understand the physical properties of Kagome materials.

In this work, we fabricated Josephson junctions of intrinsically non-superconducting, potassium-deficient $K_{1-x}V_3Sb_5$ ($x \sim 0.26 - 0.31$), and then induced superconductivity through proximity. The Josephson effect is observed in long channels up to 6 μm, and the JJs exhibit a prominent asymmetry and reversion of the magnetoresistance for the up and down magnetic field sweeps only in the superconducting state of the JJ. Moreover, the interference pattern of critical current ($I_c$) vs magnetic field shows a typical Fraunhofer-like pattern for an applied in-plane field, but an anomalous pattern with a minimum near zero-field for an applied out-of-plane field. These unusual magnetic field modulations of supercurrent with strong directional anisotropy indicates the superconducting coupling in the junction is influenced by the internal anisotropy of $K_{1-x}V_3Sb_5$ which should be related to the charge order state and the internal magnetic field. The possible spin-triplet pairing in the JJs is discussed. Finally, a non-vanishing, fast oscillation of $I_c$ with a scalloped peak-profile and excitation branches was observed, indicating the presence of edge localized supercurrent. Theoretical analysis of the band structure confirms the expectation of topological surface states in the (010)/(100) plane of $KV_3Sb_5$, corresponding to the edges of our samples. This solidifies $K_{1-x}V_3Sb_5$ as an exciting platform to study the interplay of superconductivity, quantum magnetism, and topology.

## Results

**Proximity effect induced superconductivity**. $KV_3Sb_5$ crystallizes in the hexagonal centrosymmetric space group *P*6/mmm and is composed of Kagome layers of vanadium interleaved with honeycomb layers of Sb and K (Fig. 1a). Josephson junctions with different channel lengths were fabricated on $K_{1-x}V_3Sb_5$ nanoflakes with Nb ($T_c \sim 6$ K) as the superconducting electrodes. The nanoflakes were exfoliated from the same batch of crystals used in the work of Yang *et al*(*18*), which were grown using the same method as previously described by Ortiz *et al*(*7*). It should be noted that although stoichiometric $KV_3Sb_5$ can intrinsically superconduct(*15*), the potassium deficient $K_{1-x}V_3Sb_5$ flakes can be non-superconducting. We fabricated $K_{1-x}V_3Sb_5$ devices with Au contacts to study the intrinsic properties at ultra-low temperatures. Superconductivity was observed in some Au devices with $T_c \sim 0.6$ K-0.65 K and $I_c \sim$ 13-18 μA, but was absent in others. Using electron dispersive x-ray Spectroscopy (EDS) measurements, it was revealed that there were different compositions in the superconducting Au-contacted samples compared with the non-superconducting samples; the superconducting samples were closer to the ideal 1:3:5 stoichiometry compared to the non-superconducting samples (see details in the section S3), which is consistent with the observation of intrinsic superconductivity in stoichiometric bulk $KV_3Sb_5$ with a maximum $T_c \sim 0.93$ K(*15*). The reduced $T_c$ in the deficient samples compared with stoichiometric $KV_3Sb_5$, and the disappearance of superconductivity in our $K_{1-x}V_3Sb_5$ samples with low potassium concentration, indicate that the superconductivity in $KV_3Sb_5$ is highly dependent on defect/doping. In this work, the non-superconducting $K_{1-x}V_3Sb_5$ samples with significant potassium deficiency (x = 0.26 - 0.31) were used to fabricate the Josephson junctions.

Figure 1b shows a schematic structure of a Nb/$K_{1-x}V_3Sb_5$/Nb device and one of the fabricated devices with channel length $L$ varying from 0.1 μm to 6.05 μm with a constant sample thickness of ~ 45 nm (Device #1). We first measured $R$ vs $T$ curves of both short ($L$ = 0.1 μm) and intermediate channels ($L$ = 1.88 μm) by the four-probe method to exclude the contact resistance (results shown in Fig. 1c). A sudden drop in resistance is observed at temperatures of 0.76 K and 0.68 K for channels of $L$ = 0.1 μm and $L$=1.88 μm, respectively, with the same artifactual residual resistance value from lock-in measurements (see discussion in section S1). The $I$ vs $V$ curves measured at ~20 mK show a sudden jump in voltage above $I_c$, further confirming the superconducting state at low temperature (inset of Fig. 1c).

To verify that the superconducting signal is related to the proximity effect from the superconducting electrodes, we investigated the length dependence of superconductivity by measuring the $I$ vs $V$ curves of different channel lengths at 20 mK using the two-probe method. Figure 2a shows the results measured in Device #1; the superconducting transition is observed up to 6.05 μm. The extracted critical current $I_c$ (taken from the peak location on d$V$/d$I$ vs $I$ curve) is small ($I_c$< 1μA) and different for different channel lengths, consistent with proximity induced superconductivity. Surprisingly, $I_c$ increases with increasing channel length (Fig. 2b red line), distinct from a typical JJ with a transparent interface in which $I_c$ monotonically reduces with increasing channel length (at fixed thickness and width) due to reduced coupling between the two superconducting electrodes(*51*). However, since the contact resistance of the interface is higher than the sample resistance in our device, the interface of each channel needs to be considered. We extracted the $I_cR_N$ of different JJs, where $R_N$ is the normal state resistance of the junction. As shown in Fig. 2b (purple line), $I_cR_N$ varies weakly with channel length, indicating that the $I_c$ is indeed being dominated by the interface of the SC with the $K_{1-x}V_3Sb_5$ barrier; similar $I_cR_N$ are also observed in other devices (See results of Device #2 in section S2). This again shows the supercurrent is induced by proximity effect of the superconducting electrodes.

The proximity effect induced superconductivity is also supported by many other evidences. The superconducting transition temperatures and critical current of different junctions made on the same flake (with uniform width and thickness) are different (Fig. 1c and Fig. 2), which is fundamentally distinct from an intrinsically superconducting flake with uniform width and thickness that would have constant $T_c$ and $I_c$. And we compare the magnetic field dependence of the resistance and critical current (discussed in detail later) in highly-potassium deficient Nb/$K_{1-x}V_3Sb_5$/Nb junction (Figure 3 and 4) with Au-contacted superconducting $K_{1-x}V_3Sb_5$ sample (Figure S4 in supplementary Materials). A Fraunhofer pattern was observed (Figure 4a) in the Nb-contacted junction, but disappeared in the Au-contacted superconducting sample (Figure S4), which is a strong evidence for Josephson coupling in our Nb/$K_{1-x}V_3Sb_5$/Nb junction, and against intrinsic superconductivity. Moreover, the critical field ($B_c$) of superconductivity in the Nb-contacted junction (e.g. $B_c$ ~300 mT for in-plane field and ~ 85 mT for out-of-plane field for $L$=6.05 μm) is much higher than that in the intrinsic Au-contacted superconducting $K_{1-x}V_3Sb_5$ (~100 mT for in-plane field, and ~15-20 mT for out-of-plane field). The Fraunhofer pattern also sustains up to ~ 300 mT (Figure 4a) for in-plane field. These results strongly indicate that the superconductivity in Nb-contacted junctions is coming from proximity effect of superconducting electrodes. In addition, the contribution of inhomogeneous superconductivity in the junction can also be excluded. A more comprehensive discussion of the proximity effect induced superconductivity is included in the Supplementary Materials Section S4.

While the proximity length up to 6.05 μm is quite long, much longer than the coherence length ($\xi_N$~230 nm) calculated based on bulk mobility of the barrier material (see details in Methods), similar results have been observed in JJs with a giant proximity effect when the non-superconducting bridge material in a JJ is close to superconducting itself (*i.e.* can superconduct with mild doping or temperature is above $T_c$ of barrier)(*52*, *53*), or in topological materials with topological edge/surface states(*54*, *55*). Since stoichiometric $KV_3Sb_5$ can be an intrinsic superconductor, the $K_{1-x}V_3Sb_5$ sample in the JJ may be relatively near a superconducting state and helping create the long Josephson length. This may also be the reason of weaker length dependence of superconductivity in our junction compared with typical Josephson junctions. In addition, the topological properties of $K_{1-x}V_3Sb_5$ (addressed below) may also contribute to the observed long junction lengths.

**Magnetic field sweep direction dependence of magnetoresistance**. We next study the magnetic field dependence of superconductivity in the $K_{1-x}V_3Sb_5$ JJs with both in-plane and out-of-plane magnetic field. Figure 3 shows the results of the channel with $L$=6.05 μm in Device #1. When applying the magnetic field (labeled $B_y$) in the sample plane perpendicular to the current direction (labeled $I_x$) at 20 mK, the breaking of superconductivity is observed at an in-plane critical field ($B_c$) of ~ 300 mT (Fig. 3a). In particular, the $R$ vs $B$ curve is asymmetric about zero-field for positive and negative field, but the curve reverses for up-sweep (blue line) and down-sweep (red line) of the field, as indicated by the black arrows in Fig. 3a. This property is also observed when applying an out-of-plane magnetic field ($B_z$ direction). As shown in Fig. 3b, the $R$ vs $B$ curves at 20 mK present a prominent asymmetry and reversion for up-sweep and down-sweep with an out-of-plane $B_c$ of ~ 85 mT. This flipping property of the $R$ vs $B$ curves is observed in all the measured $K_{1-x}V_3Sb_5$ JJs and channels for both in-plane and out-of-plane fields (see more data for other channels in Supplementary Materials Fig. S6). Moreover, Fig. 3b presents a prominent fast oscillation overlaid on the background of the $R$ vs $B$. Using Fast Fourier Transform analysis, the frequency of fast oscillation is extracted and found to be 1.9 $mT^{-1}$, and is not sensitive to temperature until very near the superconducting transition, as shown in Fig. 3c (additional data in Supplementary Materials Fig.

S9). This robust fast oscillation is related to the interference of supercurrent and is modulated by the magnetic flux, like a superconducting quantum interference device (SQUID), and its origin is discussed in detail below. Similar to the background of the $R$ vs $B$ curves, the fast oscillation also reverses for different field sweep directions (Fig. 3b).

**Anisotropic interference pattern for in-plane and out-of-plane magnetic field**

To further understand the magnetic field dependence of superconductivity, the magnetic field modulation of the critical supercurrent in the $K_{1-x}V_3Sb_5$ JJs is studied. Figs. 4a and 4b are the color maps of d$V$/d$I$ vs $I$ and $B$ for in-plane field and out-of-plane field respectively, measured by down-sweep of the field for the 6.05 μm JJ (see Fig. S8 in Supplementary Materials for up-sweep pattern). There are two sets of interference patterns of critical currents (bright lines) on the color maps, which may be indicative of distinct superconducting channels in the $K_{1-x}V_3Sb_5$ JJ (see detailed discussion in section S8). Since the outside pattern sustains to higher field than the inside pattern and dominates the primary interference features, we focus on the outside pattern. When applying the in-plane magnetic field, with $B_y \perp I_x$, the main interference pattern of $I_c$ vs $B$ shows a prominent peak at zero field and the oscillation of $I_c$ decays quickly with increasing field, as is expected of a Fraunhofer-like $I_c$ vs $B$ pattern for a JJ. This confirms that Josephson coupling is achieved in these devices. However, when applying an out-of-plane ($B_z$) magnetic field (Fig. 4b), the interference pattern of critical current shows an anomalous suppression near zero field, resulting in a minimum instead of a peak, at the center of the interference pattern. Under a weak magnetic field (either up or down), the $I_c$ is enhanced relative to zero-field, but is still smaller than the maximum $I_c$ with in-plane field. This anisotropic interference pattern and suppression of $I_c$ are quite unusual. In addition, a long-lived fast, SQUID-like oscillation is also clearly visible on top of the primary background in Fig. 4b. Since the superconductivity presents many properties, we discuss these properties piecewise and their related physical origins in detail below.

## Discussion

**Anisotropic internal magnetic field and spin-triplet superconductivity**

The results in Fig. 3 and Fig. 4 show the distinct magnetic field dependence of superconductivity in a $K_{1-x}V_3Sb_5$ JJ including the reversion of the $R$ vs $B$ curves when changing field sweeping direction, and anisotropic interference patterns with central peak for in-plane field, but abnormal suppression of critical current for out-of-plane field. These features are also observed in the superconducting state of other Nb-contacted $K_{1-x}V_3Sb_5$ junctions (Fig. S5 and Fig. S6 in Supplementary Materials), but disappear in Au-contacted superconducting devices (See details in Fig. S4), indicating that they are properties of the proximity-induced supercurrent in the JJs.

In $KV_3Sb_5$, no long-range magnetic order has been observed down to 0.25 K via neutron scattering (below the $T_c$ of these JJs)(*7*), however, it has presented large anomalous Hall effect, and a signal of TRS breaking with weak internal local magnetic field was detected below the CDW transition via muon spin relaxation (*22*); similar TRS breaking below CDW is also widely observed in its sister materials $CsV_3Sb_5$(*23*, *56*) and $RbV_3Sb_5$(*57*). We note that the signal of CDW transition and anomalous Hall effect also appear in our non-JJ, Au-contacted non-superconducting $K_{1-x}V_3Sb_5$ samples (See data in Fig. S3), which indicates the TRS breaking with internal magnetic field is preserved in our potassium-deficient samples. In JJs, the internal magnetic field of the barrier will influence the injection and transport of Cooper pairs(*41*), which can lead to the reversion of $R$ vs $B$ curves, as has been widely observed in JJs with barrier materials breaking TRS, such as ferromagnets(*58*), magnetic element doped barriers(*48*). The reversion of magnetoresistance is robustly observed in all $K_{1-x}V_3Sb_5$ JJs with different channel lengths, indicating it's an intrinsic property that should be induced by the internal local magnetic field in $K_{1-x}V_3Sb_5$. As a control, we also fabricated the Nb/graphene/Nb JJ and the reversion feature is absent, as expected (See data of a Nb/graphene/Nb JJ in Supplementary Materials Fig. S7).

The distinct interference patterns for in-plane and out-of-plane field in $K_{1-x}V_3Sb_5$ JJ implies that the internal field is also anisotropic, as the superconducting coupling in the junction was modulated differently when changing the direction of the external magnetic field. Recently, muon spin relaxation experiments have reported anisotropy of the internal local magnetic field between in-plane and out-of-plane directions in the sister compound $CsV_3Sb_5$(*23*). The emergence of local internal field was found to be associated with the charge order transition(*23*, *24*). So far, the low temperature order states in $KV_3Sb_5$ and $CsV_3Sb_5$ are still not clear, with several types of charge/spin orders breaking time-reversal symmetry having been proposed, including a chiral flux phase(*30*), charge/spin bond order(*33*, *34*), and spin density wave(*32*), etc. It's noteworthy that the easy plane of these orders lies in-plane (Kagome lattice), and it was also proposed that the out-of-plane magnetic field (along *c*-axis) can modulate these orders(*32*), and obvious enhancement of the magnetic field signal under out-of-plane field was detected in muon and optical experiments(*23*, *56*). The field modulatable ordered states and related internal magnetic field likely contribute to the anisotropic magnetic field response of supercurrent in JJ.

Associated with this anisotropic magnetic field dependence of the supercurrent, we discuss the origin of the suppressed $I_c$ with a minimum near zero field in the interference pattern (out-of-plane field). In general, a JJ with normal spin-singlet pairing of non-magnetic barriers displays a standard single-slit Fraunhofer interference pattern, with periodic oscillations and a central maximum. The unusual central minimum of $I_c$ normally appears in JJs with magnetic barriers where the junction has modulated phase and spin-triplet paring(*41*, *59*). A very similar suppression and minimum of $I_c$ was observed in JJs with long range spin-triplet pairing, such as the JJ of half-metal $CrO_2$ and in SF'FF'S junctions(*44*, *60*). The conversion of spin singlet to triplet pairing in JJs generally happens based on inhomogeneous magnetizations, non-collinear spin structures, or spin spiral structures, which breaks spin rotation symmetry allowing spin-flipping at the interfaces(*41*, *43*, *46*).

For the $KV_3Sb_5$ family, as discussed before, TRS breaking was reported in many works, and several possible charge/spin order states were predicted in the Kagome plane(*23*, *30*, *32–34*, *56*), which can be modulated by external out-of-plane magnetic field(*23*, *32*, *56*) and many experiments have detected enhanced magnetic field signal under this field direction(*23*, *56*). Amongst these, one widely discussed order is chiral charge order with loop current, which induces a varying local magnetic field (dependent on the hexagons vs. triangles) in Kagome lattices (*24*, *30*, *56*). Other predicted spin orders also have varying spin directions (*32*, *33*, *61*). In addition, rotation-symmetry breaking in $KV_3Sb_5$(*25*) was also reported at low temperatures in STM study. These broken symmetries combined with the spatial variation of local internal magnetic field make it possible to convert spin-singlet to spin-triplet supercurrent in $K_{1-x}V_3Sb_5$ JJs, and allow modulation of the superconducting coupling when the out-of-plane field tunes the ordered state and internal field. This also aligns with theoretical studies which found that both spin singlet and triplet pairing can appear in the $KV_3Sb_5$ family due to the complex band structure and electron correlations (particularly as a function of Fermi level which can be modulated by doping)(*28*, *62*, *63*), and that triplet pairing may be a favored state in the (relatively) weak correlation region (*28*). The combination of the observed abnormal interference pattern and these inherent properties of $K_{1-x}V_3Sb_5$ indicate that spin-triplet pairing is possible in $K_{1-x}V_3Sb_5$ JJs and more experimental effort in future, including elucidation of the ground state in the $AV_3Sb_5$ family, will help understand the superconducting behavior in $K_{1-x}V_3Sb_5$ JJs.

**Supercurrent of edge state**

We next analyze the oscillations in the interference pattern, and focus on the spatial distribution of the supercurrent and the prominent long-lived fast, SQUID-like oscillation of $I_c$ in the Fig. 4. In the junction, the $I_c$ is modulated by the magnetic flux in units of the flux quantum $\Phi_0$ with $\Phi_0 = \Delta B * L * t_{\text{eff}}$ ($L \approx 6.05$ μm). Based on the Fraunhofer-like pattern (period of $\Delta B \sim 100$ mT) for the in-plane magnetic field, an effective superconducting thickness of $t_{\text{eff}} \approx 3.4$ nm is extracted. We also performed inverse Fourier transforms of the in-plane interference pattern (see section S9), the obtained current density profile shows that supercurrent is uniformly distributed through the top ~ 5 nm of the flake, which is in quite good agreement with $t_{\text{eff}}$. These results imply that only a thin layer on the top of the flake is being proximitized, as expected; the superconducting order is not extending deeply along the *c*-axis across the layers, and is dramatically longer in the *ab*-plane. This thin proximitized layer is reasonable for $K_{1-x}V_3Sb_5$ due to its layered structure, which is expected to be electronically anisotropic ($c$ axis vs the *ab*-plane) like its sister material, $CsV_3Sb_5$, that has shown out-of-plane resistivity 600x larger than in-plane (*14*). This results in a small coherence length along the *c*-axis, just like other layered materials including $WTe_2$(*64*).

The interference pattern for the out-of-plane magnetic field, however, with fast oscillations on top of the background is not a typical Fraunhofer pattern. The SQUID-like fast oscillation maintains a period of $\Delta B \sim 0.5$ mT that is consistent with the frequency of the oscillation on the $R$ vs $B$ curves and corresponds to an effective flux-penetration area of $S_{\text{eff}} \sim 4.13$ μm$^2$ calculated from $\Phi_0 = \Delta B * S_{\text{eff}}$. This is smaller than the crystal area ($S = L * W \approx 12$ μm$^2$), which is likely caused by partial penetration of the flux, as was seen with $MoTe_2$(*65*). The most important feature of the fast oscillation is that it is non-vanishing, and survives to higher field than the main interference pattern, as shown in Fig. 4b. This long-lived, fast oscillation cannot be induced by the bulk Josephson current. Moreover, it presents a scalloped peak profile with clear excitation branches trailing from the scalloped boundary and prominent jumps between oscillation branches are observed (Fig. 4c). This scalloped profile and excitation branches of a non-vanishing fast mode are important signatures of non-bulk, spatially localized states (*i.e.* not uniformly distributed supercurrent) like surface or edge states(*39*, *65*, *66*). For an enclosed area bounded by the sides of the sample, the requirement of flux quantization results in a sawtooth profile of the surface Fermi velocity, which contributes to the scalloped profile and excitation branches, as observed in the edge-state of $MoTe_2$(*65*). We also performed inverse Fraunhofer transform for the out-of-plane field interference pattern to extract the supercurrent density distribution along the width direction, and although the quantitative analysis has some error due to the complicated $I_c$ vs $B$ pattern, the existence of two clear peaks in the current profile qualitatively indicates the presence of edge supercurrent (see details in Supplementary Materials Fig. S11).

To examine the origin of this edge supercurrent, we further performed density functional theory (DFT) calculations of the electronic band structure of $KV_3Sb_5$. Previous work on both the isostructural, isoelectronic compound $CsV_3Sb_5$ as well as $KV_3Sb_5$ have predicted $\mathbb{Z}_2$-protected topological surface Dirac crossings just above the Fermi level(*14*, *15*). Here we calculated the spectral density of bulk (upper panel) and surface states (lower panel) particularly on the (100)/(010) planes (*ac*/*bc* planes), including states propagating along the [001] directions, as results shown in Fig. 4d. Although the large number of surface states prevents us from attributing the surface conductivity to a single surface band, our calculations show there are prominent surface states near the Fermi level along the $\bar{X}-\bar{M}$ line, and the Fermi level of $K_{1-x}V_3Sb_5$ (slightly below the predicted level, as shown by previous ARPES measurements(*18*)) cuts the surface states near the $\bar{X}$ point (corresponding to the edges in our thin flake JJ device). These surface states should contribute to the edge supercurrent shown as SQUID-like fast oscillations in the $K_{1-x}V_3Sb_5$ JJ. See the methods and Supplementary Materials section S10 for more details of surface states. It's should be noted that the surface states in the *ab*-plane were not detected when applying in-plane field along the *y*-axis, because only the top layers of the $K_{1-x}V_3Sb_5$ samples were proximitized; coupling between the top and bottom surface states to result in the SQUID-like pattern was not possible, in analogy to what was seen with the hinge-states in $Al_2O_3$ back-filled $WTe_2$ devices(*67*). Detecting the *ab*-plane surface states is an area for future work and will require very thin flakes (sub 10 nm) or vertical JJ architectures.

In summary, we fabricated JJs of the topological Kagome metal, $K_{1-x}V_3Sb_5$, and observed Josephson coupling through long distance; asymmetry and reversion of $R$ vs $B$ as well as a zero-field minimum in the $I_c$ vs $B$ pattern which indicate anisotropic internal magnetism and possible spin-triplet superconductivity; and also the signal of supercurrent of the predicted topological surface states in $K_{1-x}V_3Sb_5$. These observations in one system open the door for exploring the combination of superconductivity with quantum magnetism and topology. Fundamentally, there is a variety of opportunities for future theoretical and experimental work to excavate unconventional superconductivity based on Josephson junctions of Kagome materials.

## Methods

### Device fabrication and measurement

Thin flakes of $KV_3Sb_5$ were exfoliated onto an $SiO_2$/Si substrate, followed by standard electron beam lithography (EBL) and sputtering processes to fabricate the Josephson Junction (JJ) devices with Niobium (~ 40 nm) as the superconducting electrodes.

The JJ's were measured in a Bluefors dilution refrigerator with base temperature reaching ~ 15 mK. The resistance of device was measured by applying different AC current excitations with low frequencies (below 30 Hz) and Zurich lock-in amplifiers. The differential resistance ($dV/dI$) was measured by adding an external DC source onto the small AC source of lock-in.

### Density functional theory (DFT) calculations

The electronic structure of $KV_3Sb_5$ was simulated in VASP v5.4.4(*68–70*) using projector-augmented wave (PAW) potentials(*71*, *72*) with identical parameters to a previous work on $CsV_3Sb_5$(*14*). Calculations employed the PBE functional(*73*) with D3 dispersion corrections(*74*), a Γ-centered 11×11×5 k-mesh, a plane wave energy cutoff of 500 eV, and the recommended PAW potentials for v5.2. No magnetic moments were incorporated in the simulation. *a* and *c* lattice parameters of the relaxed cell were 5.42 Å and 8.92 Å; in good agreement with experimental lattice parameters of 5.48 Å and 8.95 Å. Similar PBE-functional calculations have been shown to accurately recreate the slab electronic structure measured in ARPES experiments for both $CsV_3Sb_5$(*14*) and $KV_3Sb_5$(*18*). Wannier90 was used to generate an empirical tight-binding model starting from initial projectors corresponding to valence orbitals (*V p, d*; *Sb p*; with a frozen inner fitting window $E_F \pm 2$ eV, and an outer window of $E_F$ - 6 eV to $E_F$ + 5 eV)(*75*). Surface density of states calculations were performed using the method of Sancho and coworkers(*76*) as implemented in WannierTools(*77*).

## Calculation of coherence length

For $K_{1-x}V_3Sb_5$, according to equation $\mu = el_e/\hbar k_F$ with mobility $\mu \sim 0.2 m^2/Vs$ (*18*) and $k_F \sim 0.032$ Å$^{-1}$, the mean free path of $l_e \sim 42$ nm is extracted. Since $l_e$ is smaller than the coherence length of Nb ($\xi_S = \frac{\hbar V_F}{1.76\pi k_B T_c} \approx 64$ nm, $T_c \approx 6.2$ K, $V_F \approx 3 \times 10^5 m/s$)(*78*), the superconducting coherence length of spin-singlet pairing in non-magnetic materials(*52*, *64*) or spin-triplet pairs in a magnetic system(*48*) can be calculated by $\xi_N = \sqrt{\hbar v_F l_e / 6\pi k_B T_c}$ (Fermi velocity $v_F \sim 3.77 \times 10^5 m/s$, $T_c \sim 0.78$ K), which is about 230 nm in $K_{1-x}V_3Sb_5$.

## Supplementary Materials

**This PDF file includes:**

Sections S1 to S10

Figs. S1 to S14

## Acknowledgments

We thank Fang Gao and Jae-Chun Jeon for the support during device fabrication, and thank Gil-Ho Lee and Vic K. T. Law and for valuable discussions. **Funding:** M.N.A acknowledges that this research was principally supported by the Alexander von Humboldt Foundation Sofia Kovalevskaja Award, the German Federal Ministry of Education and Research's MINERVA ARCHES Award, and the Max Planck Society. Y.W acknowledges the support from NWO Talent Programme Veni financed by the Dutch Research Council (NWO), project No. VI.Veni.212.146. S. D.W., B. R. O., and S. M. L. T. acknowledge support from the University of California Santa Barbara Quantum Foundry, funded by the National Science Foundation (NSF DMR-1906325). Research reported here also made use of shared facilities of the UCSB MRSEC (NSF DMR-1720256). B. R. O. also acknowledges support from the California NanoSystems Institute through the Elings fellowship program. S.M. L. T. has been supported by the National Science Foundation Graduate Research Fellowship Program under Grant No. DGE-1650114. Any opinions, findings, and conclusions or recommendations expressed in this material are those of the authors and do not necessarily reflect the views of the National Science Foundation. D.L. acknowledges support from the Alexander von Humboldt Foundation. S.S.P.P. acknowledges the European Research Council (ERC) under the European Union's Horizon 2020 research and innovation programme (grant agreement no. 670166), Deutsche Forschungsgemeinschaft (DFG, German Research Foundation)—project number 314790414, and Alexander von Humboldt Foundation in the framework of the Alexander von Humboldt Professorship endowed by the Federal Ministry of Education and Research. T.M. acknowledges the David and Lucile Packard Foundation and the Johns Hopkins University Catalyst Award. E.S.T. acknowledges support of NSF DMR 1555340.

**Author contributions:** Y.W. and M.N.A. conceived and designed the study. B.R.O. grew the samples. Y.W. and S.-Y.Y. fabricated the devices. Y.W., P.K.S. and S.-Y.Y. performed the transport measurement. Y.W, H.W, S.-Y. Y., C. G. and D.L carried out the data analysis. S.M.L.T. carried out DFT calculations and theoretical analysis. A.K.S. performed the EDX measurements. E.S.T., T.M., S.D.W., and M.N.A are the Principal Investigators. All authors contributed to the preparation of manuscript.

**Competing interests:** The authors declare that they have no competing interests.

**Data and materials availability:** All data needed to evaluate the conclusions in the paper are present in the paper and/or the Supplementary Materials.

**Correspondence and requests for materials** should be addressed to Y. W. and M.N.A.

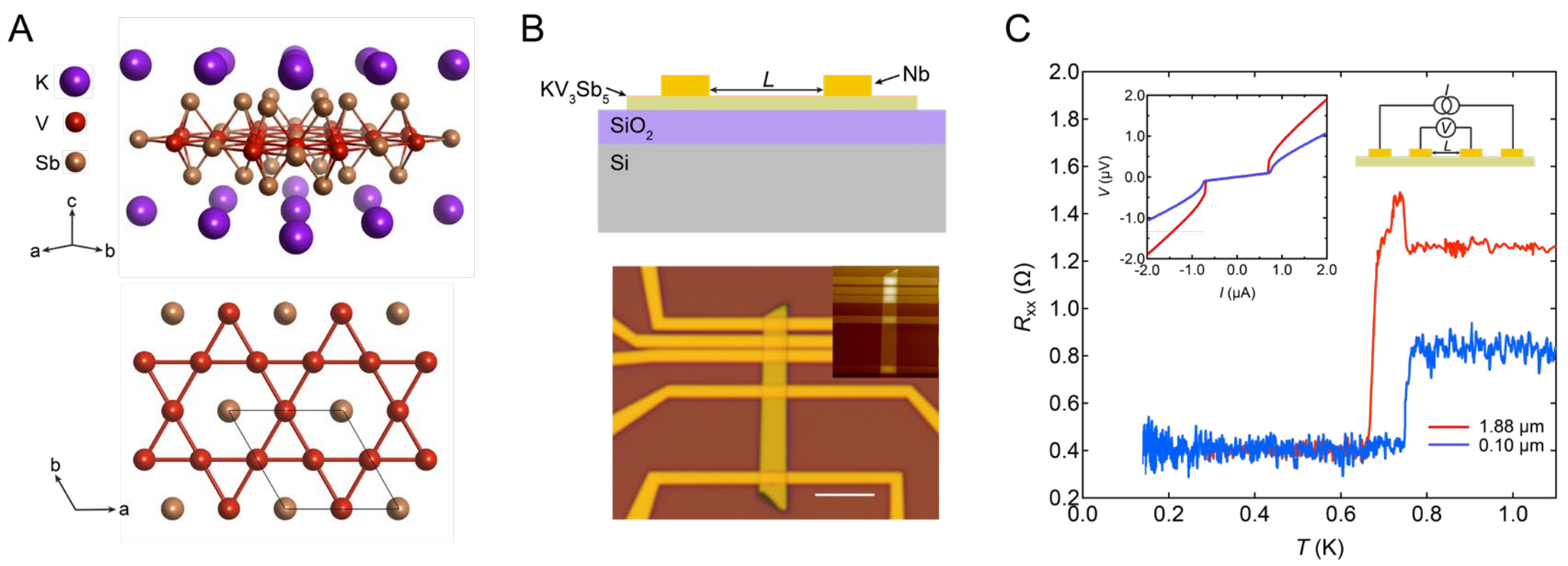

**Fig. 1. Crystal structure and Josephson junction of $KV_3Sb_5$.** **(A)** Top: Crystal structure of $KV_3Sb_5$, Bottom: Projection along *c*-axis showing the Kagome net of vanadium. **(B)** The top panel is side-view schematic of $K_{1-x}V_3Sb_5$ JJ, the bottom panel is the optical and AFM (inset) image of one fabricated JJ device (Device #1) of $K_{1-x}V_3Sb_5$ thin flake (~ 45 nm), the scale bar is 5 μm. **(C)** Superconducting transition of Device #1 measured by four-probe method. The insets are schematic of measurement circuit (top right) and typical voltage versus current curves measured at 20 mK (top left).

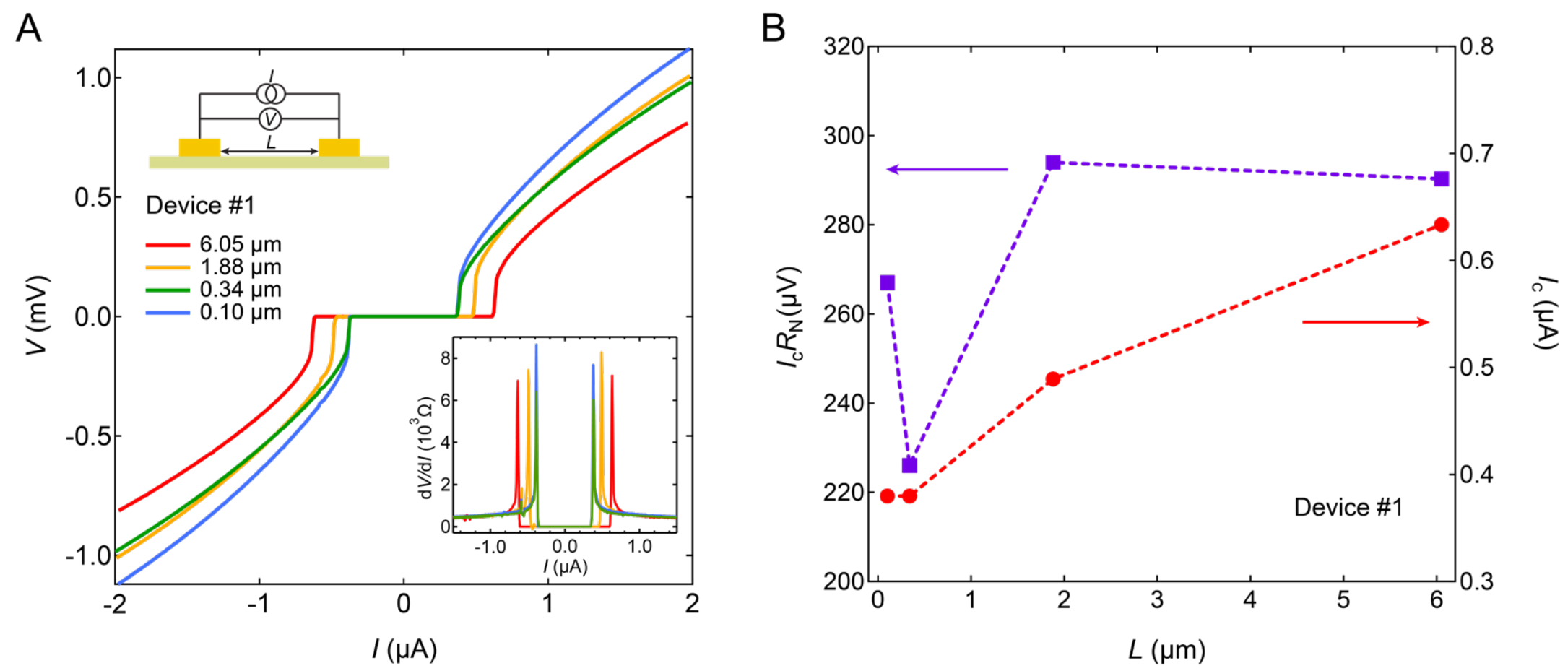

**Fig. 2. Length dependence of Josephson current. (A)** Voltage versus current ($V$-$I$) curves for different channel lengths measured by two-probe method at 20 mK. The insets are corresponding differential resistance (down right) and schematic of measurement circuit (top left). **(B)** Length dependent critical currents $I_c$ and $I_cR_N$ extracted from the $V$-$I$ curves in (**A**).

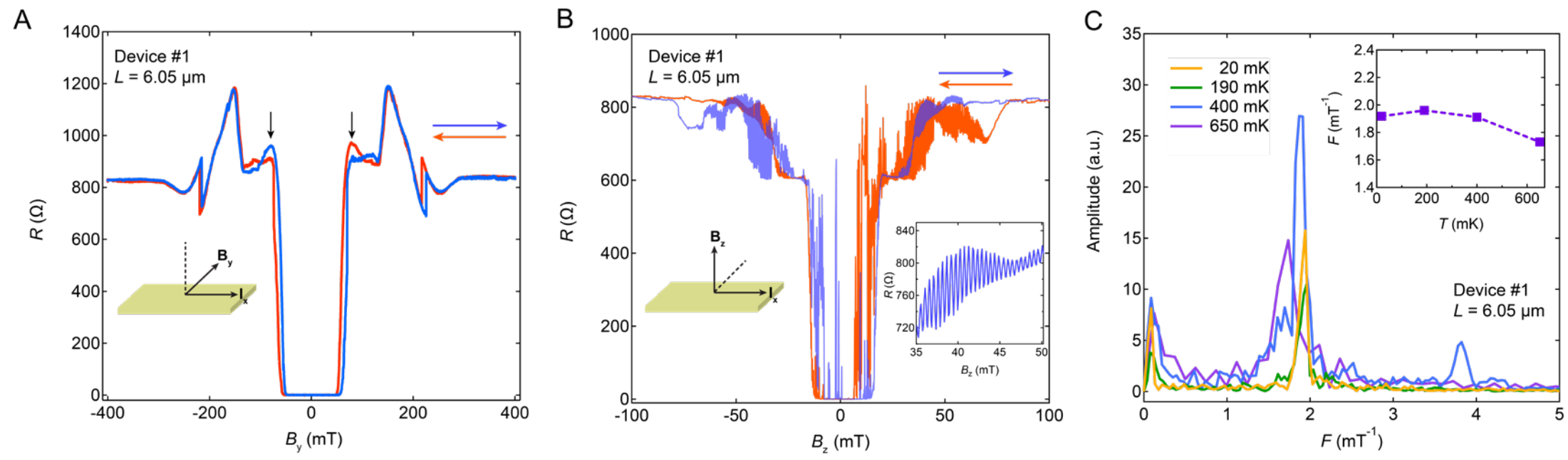

**Fig. 3. Magnetic field dependent resistance of Josephson junction with *L*=6.05 μm. (A)** *R* vs *B* with magnetic field applied in-plane but orthogonal to the current. Red (blue) line and arrows denote field down (up) sweeping. The two black arrows point out the reversion of resistance for different sweep directions. The inset denotes magnetic field and current directions. **(B)** *R* vs *B* of the same device with magnetic field applied out of plane and orthogonal to the current. The left inset denotes magnetic field and current directions and the right inset is the enlarged plot of fast oscillation. **(C)** Fast-Fourier transform (FFT) of fast oscillation in (**B**) at different temperatures, the inset is temperature dependent oscillation frequency obtained in the main panel of (**C**).

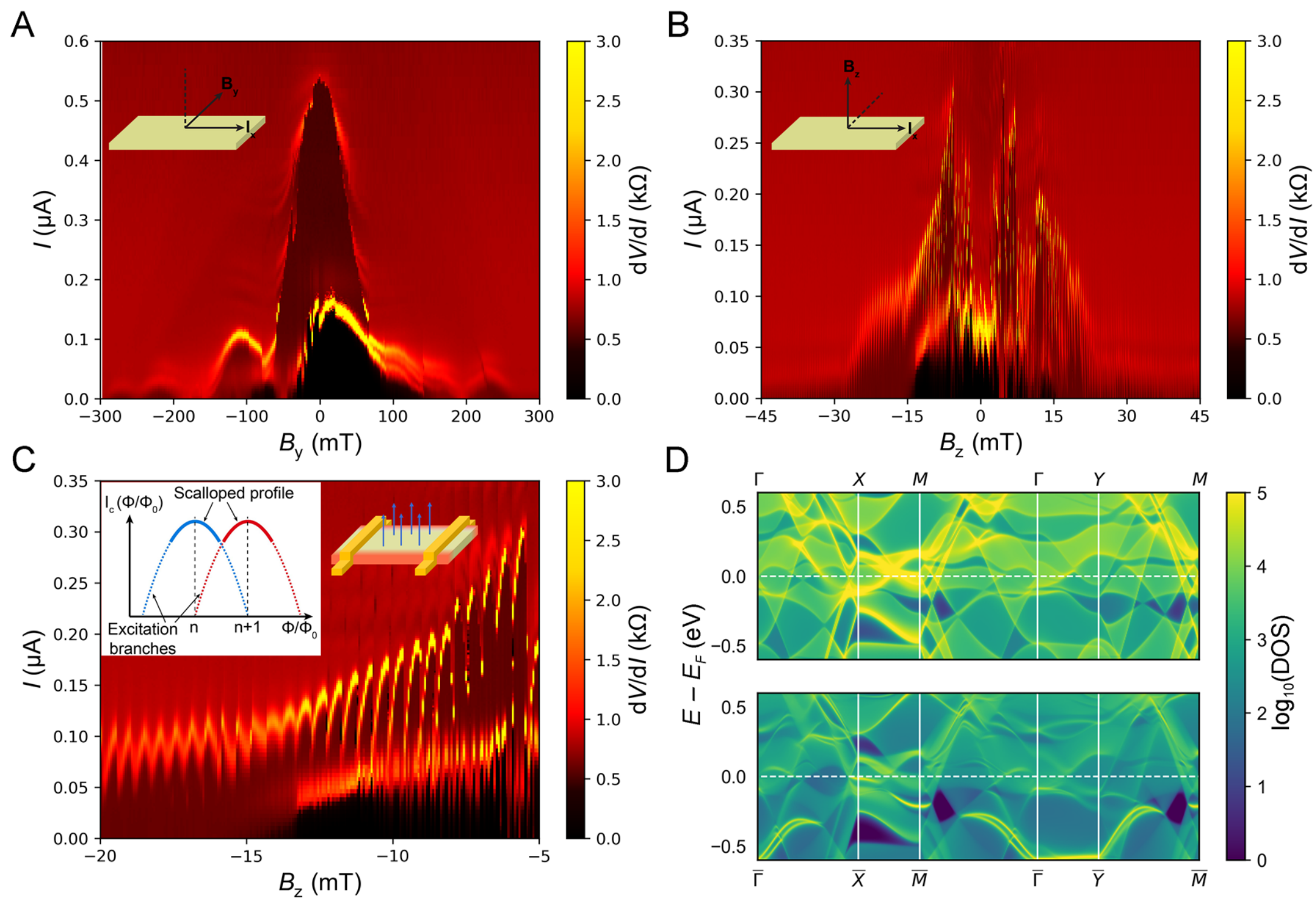

**Fig. 4. Interference patterns and localized Josephson current.** **(A)** and **(B)** are color maps of d$V$/d$I$ versus current and magnetic field measured in Device #1 at 20 mK with applying in-plane magnetic field and out-of-plane magnetic field, respectively. The field is changing from positive to negative during the measurement. The insets denote magnetic field and current directions. **(C)** Main panel: the enlarged plots of the fast oscillation in (**B**) with applying out-of-plane field. The right inset is a schematic of the edge supercurrent in the JJ, the edges are indicated by red color. The left inset is a schematic current profile of an edge state. **(D)** Top: Bulk spectral density of states for $KV_3Sb_5$ on the (010)/(100) surfaces. Bottom: Surface spectral density of states for the (010)/(100) surfaces. Several bright bands are present in the bottom panel but not the top panel; these are the surface states.